\documentclass[screen,nonacm]{acmart}
\usepackage{graphicx} 

\usepackage[altpo]{backnaur}  

\usepackage{listings}

\definecolor{notlobstruct}{HTML}{1B3A5C}
\definecolor{notlobsigil}{HTML}{5B2A6E}
\definecolor{notlobbullet}{HTML}{6E5B1B}
\definecolor{notlobprose}{HTML}{2E4D6B}

\newenvironment{lobprose}
  {\itshape\color{notlobprose}\small\par\vspace{0.3em}}
  {\par\vspace{0.3em}\normalfont\normalcolor}

\lstdefinelanguage{notlob}{
  literate=
    {~property}{{\textcolor{notlobsigil}{\textasciitilde{}property}}}9
    {~example}{{\textcolor{notlobsigil}{\textasciitilde{}example}}}8
    {~test}{{\textcolor{notlobsigil}{\textasciitilde{}test}}}5
    {~run}{{\textcolor{notlobsigil}{\textasciitilde{}run}}}4
    {---}{{\textcolor{notlobstruct}{---}}}3
    {*}{{\textcolor{notlobbullet}{*}}}1,
  basicstyle=\ttfamily\small,
  columns=fullflexible,
  keepspaces=true,
  breaklines=true,
  showstringspaces=false,
  morecomment=[l]{},   
}

\newcommand{\concept}{\textit}

\newcommand{\sigil}[1]{\texttt{\char`\~#1}}

\newcommand{\synt}[1]{\textit{#1}}

\colorlet{punct}{red!60!black}
\definecolor{background}{HTML}{EEEEEE}
\definecolor{delim}{RGB}{20,105,176}
\colorlet{numb}{magenta!60!black}

\lstdefinelanguage{json}{
    basicstyle=\normalfont\ttfamily,
    numberstyle=\scriptsize,
    stepnumber=1,
    numbersep=8pt,
    showstringspaces=false,
    breaklines=true,
    literate=
     *{0}{{{\color{numb}0}}}{1}
      {1}{{{\color{numb}1}}}{1}
      {2}{{{\color{numb}2}}}{1}
      {3}{{{\color{numb}3}}}{1}
      {4}{{{\color{numb}4}}}{1}
      {5}{{{\color{numb}5}}}{1}
      {6}{{{\color{numb}6}}}{1}
      {7}{{{\color{numb}7}}}{1}
      {8}{{{\color{numb}8}}}{1}
      {9}{{{\color{numb}9}}}{1}
      {:}{{{\color{punct}{:}}}}{1}
      {,}{{{\color{punct}{,}}}}{1}
      {\{}{{{\color{delim}{\{}}}}{1}
      {\}}{{{\color{delim}{\}}}}}{1}
      {[}{{{\color{delim}{[}}}}{1}
      {]}{{{\color{delim}{]}}}}{1},
}

\usepackage{multicol}

\title{A Literate Programming Environment for Human and Machine Agents}

\author{Adam T. Burke}
\orcid{0000-0003-4407-2199}
\email{at.burke@qut.edu.au}
\affiliation{%
  \institution{Queensland University of Technology}
  \city{Brisbane}
  \state{Queensland}
  \country{Australia}
}

\usepackage{fancyhdr}
\AtBeginDocument{%
    \addtolength{\footskip}{2.0\baselineskip}%
    \fancyfoot[L]{\textit{\textbf{Preprint}}}%
}

\date{August 2026}

\begin{document}

\begin{abstract}
This paper introduces an environment for constructing literate programs in concert with language-aware machine agents. This environment includes a grammar for executable program essays, a parser that treats names as first-class objects, an internal name-graph which relates prose, names and executable artifacts, and a binding mechanism for existing languages and testing toolsets. This supports co-location of code with its most relevant natural language and structured data context, making better use of Large Language Model (LLM) context windows. It also provides LLM coding agents with a toolset more analogous to the symbol-aware search and usage information available in human programmer-facing Integrated Development Environments (IDEs). We describe a working implementation with bindings to three established programming languages, and several example programs.
\end{abstract}


\received{2026}

\maketitle

\section{Introduction}

We are in a technological moment when the use of Large Language Models (LLMs)~\cite{vaswani_attention_2017} has triggered major changes in programming practice. This ranges from the authoring of small scripts by programming novices, to analysis of codebases for security flaws, to the construction and extension of large software systems by experienced teams of software engineers.

As language-aware AI coding tools are changing programming, our programming languages and environments should change as well. This paper introduces a new programming language and environment informed by three observations. Firstly, and unsurprisingly, natural language text dominates the specification and design organization of programs developed with language-aware AI tooling. Secondly, machine checkable computational artifacts, which constrain the operation and execution of programs, retain or increase their value by acting as an automated feedback surface for coding agents. Thirdly, the influence of the context window is dominant in the current generation of language-aware AI.

In the transformer architecture, a very large language corpus is used to train a model, then interrogated at inference time by next token prediction. No change is made to the model during basic inference, for example, a session where a programmer works with a coding agent. The model itself does not learn. Any memory or continuity in such a session comes from the re-injection of symbols into the context window from chat history or other text artifacts.

Peter Naur has an influential argument that programming is theory building~\cite{naur_programming_1985,cockburn_agile_2006}. Crucially, Naur argues that the theory being built lives primarily in the programmer's head, not in program artifacts such as source code. From this point of view, when using coding agents, there is no way for the model itself to learn this theory, so the responsibility for building and carrying the theory remains entirely with the programmer. 

As Dominic Fox points out~\cite{fox_holding_2026}, Donald Knuth's literate programming~\cite{knuth_literate_1984} is a competing school of thought. It holds that a theory of a program can not only be conveyed in written artifacts, but that well-composed natural language is a highly effective vehicle for doing so. Knuth accordingly created tools and a style of writing code and essay-style prose together, an idea that has influenced much scholarship and language tooling since, without really centring software development on prose as Knuth envisioned. Recent research explores the sympathies between language-competent agents and literate programming~\cite{zhang_renaissance_2024,shi_natural_2025}, even declaring it a literate programming renaissance~\cite{zhang_renaissance_2024}.

``The hottest new programming language is English", according to Andrej Karpathy in 2023~\cite{karpathy_hottest_2023},  though this observation was as much fashion journalism as design intent. With this ``literate turn" in programming history, we should reach for that genre of writing which is used to explore and advance an idea: the essay. Like Knuth, we should do so without abandoning the productive mechanics of programming language design. Unlike Knuth, both humans and computers will be readers of the essays and the executable code. 

This paper presents a literate programming environment for human and machine agents. This includes:

\begin{itemize}
    \item a machine interpretable language, comprising both prose and executable elements; 
    \item embedded code fragments, using existing programming languages, and their rich libraries;
    \item an internal data structure relating both concept names and executable symbols, called the name-graph;
    \item tools which expose name-graph relationships, providing IDE-like support for agents and other tools; and
    \item example programs.
\end{itemize}

The three observations motivate this design. The dominance of natural language text leads to a language with both prose and executable elements. The usefulness of machine checkable artifacts leads to keywords for property and unit tests. The importance of the context window re-motivates literate programming's co-location of prose, code, and validation in files, and navigation using the name-graph. 
The programming language and environment is named notlob, after a section in Monty Python's parrot sketch where a dissimulating shopkeeper makes a false but easily checkable claim about palindromes and the town of Bolton. 

This paper focuses on design and example programs of the language and tooling environment; experimental evaluation is left to future work. The remainder of this paper is organised as follows. Section~\ref{sec:language} introduces the notlob grammar, semantics and name-graph. Section~\ref{sec:impl} introduces a reference implementation of the language and environment. Section~\ref{sec:programs} shares example programs and projects. Section~\ref{sec:related-work} reviews related work, and Section~\ref{sec:conclusion} concludes.

\section{Design}
\label{sec:language}

\newcommand{\egfigsize}{basicstyle=\ttfamily\footnotesize}

\begin{figure}[t]
\begin{minipage}[t]{0.48\textwidth}
\begin{lstlisting}[language=notlob,basicstyle=\ttfamily\footnotesize]
#Fibonacci
\end{lstlisting}
\begin{lobprose}
The Fibonacci sequence: each number is the sum of the two before it,
starting from 0 and 1.
\end{lobprose}
\begin{lstlisting}[language=notlob,basicstyle=\ttfamily\footnotesize]
    fib :: Int -> Int
    fib 0 = 0
    fib 1 = 1
    fib n = fib (n - 1) + fib (n - 2)
\end{lstlisting}
\begin{lobprose}
The defining recurrence, checked here as a property
rather than assumed for a handful of examples.
\end{lobprose}
\end{minipage}
\hfill
\begin{minipage}[t]{0.48\textwidth}
\begin{lstlisting}[language=notlob, basicstyle=\ttfamily\footnotesize]
~property
    prop_recurrence :: Int -> Bool
    prop_recurrence n =
        let k = abs n `mod` 20 + 2
        in fib k == fib (k - 1) + fib (k - 2)

---
#Tests

##base cases
    fib 0 == 0
    fib 1 == 1

##known values
    fib 10 == 55
\end{lstlisting}
\end{minipage}
\caption{A notlob program which calculates the Fibonacci sequence, using the Haskell binding.}
\label{fig:fibeg}
\Description{Code listing.}
\end{figure}


\newcommand{\grammarsize}{\scriptsize}

\begin{figure}[t] 
  \centering
    \grammarsize


\begin{bnf*}
\bnfprod{start}
    {\bnfpn{module}} \\
\bnfprod{module}
    {\bnfpn{MOD\_HEAD} \bnfsp \bnfpn{body} \bnfsp [\,\bnfpn{post\_text}\,]} \\
\bnfprod{body}
    {\{\,\bnfpn{body\_item}\,\}} \\
\bnfprod{body\_item}
    {\bnfpn{subheading}} \\
\bnfmore{\bnfor \bnfpn{code\_block}} \\
\bnfmore{\bnfor \bnfpn{claim}} \\
\bnfmore{\bnfor \bnfpn{prose\_block}} \\
\bnfmore{\bnfor \bnfpn{bullet\_block}} \\
\bnfmore{\bnfor \bnfpn{BLANK}} \\
\bnfprod{subheading}
    {\bnfpn{SUBHEAD} \bnfsp \{\,\bnfpn{\_sub\_item}\,\}} \\
\bnfprod{\_sub\_item}
    {\bnfpn{code\_block}} \\
\bnfmore{\bnfor \bnfpn{claim}} \\
\bnfmore{\bnfor \bnfpn{prose\_block}} \\
\bnfmore{\bnfor \bnfpn{bullet\_block}} \\
\bnfmore{\bnfor \bnfpn{BLANK}} \\
\bnfprod{code\_block}
    {\bnfpn{INDENTED\_LINE} \bnfsp \{\,\bnfpn{\_body\_line}\,\}} \\
\bnfprod{claim}
    {\bnfpn{SIGIL} \bnfsp \bnfpn{\_body\_line}\textsuperscript{+}} \\
\bnfprod{prose\_block}
    {\bnfpn{prose\_line}\textsuperscript{+}} \\
\bnfprod{prose\_line}
    {(\bnfpn{LINE\_START\_TEXT} \bnfor \bnfpn{PROSE\_TEXT} \bnfor \bnfpn{REF})\textsuperscript{+} \bnfsp \bnfpn{BLANK}} \\
\bnfprod{bullet\_block}
    {\bnfpn{BULLET}\textsuperscript{+}} \\
\bnfprod{\_body\_line}
    {\bnfpn{INDENTED\_LINE}} \\
\bnfmore{\bnfor \bnfpn{BLANK}} \\
\bnfprod{post\_text}
    {\bnfpn{SEPARATOR} \bnfsp \{\,(\bnfpn{BLANK} \bnfor \bnfpn{post\_section})\,\}} \\
\bnfprod{post\_section}
    {\bnfpn{tests\_section}} \\
\bnfmore{\bnfor \bnfpn{binding\_section}} \\
\bnfmore{\bnfor \bnfpn{references\_section}} \\
\bnfmore{\bnfor \bnfpn{appendix\_section}} \\
\bnfprod{tests\_section}
    {\bnfpn{TESTS\_HEAD} \bnfsp \{\,(\bnfpn{BLANK} \bnfor \bnfpn{test\_item} \bnfor \bnfpn{prose\_block})\,\}} \\
\bnfprod{test\_item}
    {\bnfpn{test\_group}} \\
\bnfmore{\bnfor \bnfpn{INDENTED\_LINE}} \\
\bnfprod{test\_group}
    {\bnfpn{SUBHEAD} \bnfsp \{\,(\bnfpn{INDENTED\_LINE} \bnfor \bnfpn{BLANK} \bnfor \bnfpn{prose\_block} \bnfor \bnfpn{named\_test})\,\}} \\
\bnfprod{named\_test}
    {\bnfpn{TEST\_SIGIL} \bnfsp \bnfpn{\_body\_line}\textsuperscript{+}} \\
\bnfprod{binding\_section}
    {\bnfpn{BINDING\_HEAD} \bnfsp \{\,((\bnfpn{INDENT} \bnfsp \bnfpn{bind\_detail\_decl}) \bnfor \bnfpn{BLANK})\,\}} \\
\bnfprod{bind\_detail\_decl}
    {\bnfpn{LANGUAGE\_DECL}} \\
\bnfmore{\bnfor \bnfpn{EXTERNAL\_DECL}} \\
\bnfmore{\bnfor \bnfpn{ON\_BUILD\_DECL}} \\
\bnfmore{\bnfor \bnfpn{KEEP\_SRC\_DECL}} \\
\bnfprod{references\_section}
    {\bnfpn{REFERENCES\_HEAD} \bnfsp \{\,(\bnfpn{INDENTED\_LINE} \bnfor \bnfpn{BLANK})\,\}} \\
\bnfprod{appendix\_section}
    {\bnfpn{APPENDIX\_HEAD} \bnfsp \{\,\bnfpn{body\_item}\,\}}
\end{bnf*}

\caption{Notlob grammar - Productions.}
\label{fig:notlob_grammar_productions}
\Description{Grammar listing.}
\end{figure}



\begin{figure}[t]
    \centering
    \grammarsize

\begin{bnf*}
\bnfprod{LINE\_CHAR}
    {\bnftd{any character other than newline}} \\
\bnfprod{REST\_OF\_LINE}
    {\{\,\bnfpn{LINE\_CHAR}\,\} \bnfsp \bnfpn{NewLine}} \\
\bnfprod{MOD\_HEAD}
    {\bnfts{\#} \bnfsp \bnfpn{NonHashLineChar} \bnfsp \bnfpn{REST\_OF\_LINE}} \\
\bnfprod{SUBHEAD}
    {\bnfts{\#\#} \bnfsp \bnfpn{REST\_OF\_LINE}} \\
\bnfprod{SIGIL}
    {\bnfts{\textasciitilde{}example} \bnfsp \bnfpn{NewLine}} \\
\bnfmore{\bnfor \bnfts{\textasciitilde{}run} \bnfsp \bnfpn{NewLine}} \\
\bnfmore{\bnfor \bnfts{\textasciitilde{}run on{-}load} \bnfsp \bnfpn{NewLine}} \\
\bnfmore{\bnfor \bnfts{\textasciitilde{}run on{-}invocation} \bnfsp \bnfpn{NewLine}} \\
\bnfmore{\bnfor \bnfts{\textasciitilde{}property} \bnfsp \bnfpn{NewLine}} \\
\bnfmore{\bnfor \bnfts{\textasciitilde{}property } \bnfsp \bnfpn{REST\_OF\_LINE}} \\
\bnfprod{TEST\_SIGIL}
    {\bnfts{\textasciitilde{}test } \bnfsp \bnfpn{LINE\_CHAR} \bnfsp \bnfpn{REST\_OF\_LINE}} \\
\bnfprod{LANGUAGE\_DECL}
    {\bnfts{\textasciitilde{}language } \bnfsp \bnfpn{LINE\_CHAR} \bnfsp \bnfpn{REST\_OF\_LINE}} \\
\bnfprod{EXTERNAL\_DECL}
    {\bnfts{\textasciitilde{}external } \bnfsp \bnfpn{LINE\_CHAR} \bnfsp \bnfpn{REST\_OF\_LINE}} \\
\bnfprod{ON\_BUILD\_DECL}
    {\bnfts{\textasciitilde{}on{-}build } \bnfsp \bnfpn{LINE\_CHAR} \bnfsp \bnfpn{REST\_OF\_LINE}} \\
\bnfprod{KEEP\_SRC\_DECL}
    {\bnfts{\textasciitilde{}keep{-}generated{-}src} \bnfsp \bnfpn{REST\_OF\_LINE}} \\
\bnfprod{SEPARATOR}
    {\bnfts{{-}{-}{-}} \bnfsp \bnfpn{NewLine}} \\
\bnfprod{TESTS\_HEAD}
    {\bnfts{\#Tests} \bnfsp \bnfpn{NewLine}} \\
\bnfprod{BINDING\_HEAD}
    {\bnfts{\#Binding} \bnfsp \bnfpn{NewLine}} \\
\bnfprod{REFERENCES\_HEAD}
    {\bnfts{\#References} \bnfsp \bnfpn{NewLine}} \\
\bnfprod{APPENDIX\_HEAD}
    {\bnfts{\#Appendix} \bnfsp \bnfpn{REST\_OF\_LINE}} \\
\bnfprod{INDENT}
    {(\bnfpn{Space} \bnfor \bnfpn{Tab})\textsuperscript{+}} \\
\bnfprod{INDENTED\_LINE}
    {\bnfpn{INDENT} \bnfsp \bnfpn{REST\_OF\_LINE}} \\
\bnfprod{BLANK}
    {\bnfpn{NewLine}} \\
\bnfprod{BULLET}
    {\bnfts{*} \bnfsp (\bnfpn{NewLine} \bnfor ((\bnfpn{Space} \bnfor \bnfpn{Tab}) \bnfsp \bnfpn{REST\_OF\_LINE}))} \\
\bnfprod{REF}
    {(\bnfts{\#} \bnfor \bnfts{\#\#}) \bnfsp \bnfpn{UpperLetter} \bnfsp \{\,\bnfpn{WordChar}\,\} \bnfsp \{\,(\bnfpn{Space} \bnfsp \bnfpn{UpperLetter} \bnfsp \{\,\bnfpn{WordChar}\,\})\,\}} \\
\bnfprod{LINE\_START\_TEXT}
    {\bnfpn{ProseInitial} \bnfsp \{\,\bnfpn{ProseTail}\,\}} \\
\bnfprod{PROSE\_TEXT}
    {\bnfpn{ProseTail}\textsuperscript{+}} \\
\bnfprod{NonHashLineChar}
    {\bnfpn{LINE\_CHAR} - \bnfts{\#}} \\
\bnfprod{UpperLetter}
    {\bnftd{an uppercase letter}} \\
\bnfprod{WordChar}
    {\bnftd{a Unicode letter, digit, or underscore}} \\
\bnfprod{Space}
    {\bnftd{a space character}} \\
\bnfprod{Tab}
    {\bnftd{the tab character}} \\
\bnfprod{NewLine}
    {\bnftd{the newline character}}
\end{bnf*}


  \caption{Notlob grammar - Terminals.}
  \label{fig:notlob_grammar_terminals}
  \Description{Grammar listing.}
\end{figure}

The notlob language consists of structured prose with embedded executable blocks, demarcated by headings. The macro structure of a file is modelled on a technical report. The source is divided into a main body and supplementary text. The body is intended for those functions and data structures used at runtime, including the underlying concepts. The supplemental text, after a text literal evoking a horizontal break, is for tests, appendices, and references, including code and heading imports. 
Where some approaches for programming with LLMs use specifications placed upstream of a pipeline which produces code, notlob co-locates prose, code, and checks in a single source artifact.

A small example notlob program for calculating the Fibonacci sequence is in Figure~\ref{fig:fibeg}. This main section demonstrates the heading, introductory text, and function declaration. A property declaration indicates an executable property test. After the break characters, unit tests give statements of Fibonacci facts, similar to an appendix of historical chronology or other data tables. This example uses a Haskell binding for the executable layer.

\subsection{Language Syntax}
\label{sec:syntax}

In general the gross structure supports putting the explanation, execution and supporting checks together in a single file. The necessary context for human and machine agents can often be found immediately adjacent in the same file read.

An EBNF grammar for notlob is in Figures~\ref{fig:notlob_grammar_productions}, for productions, and~\ref{fig:notlob_grammar_terminals}, for terminals. Title and section headers use a Markdown-style \concept{\#} prefix~\cite{gruber_markdown_2004}. Inline references to headings, and so contexts, uses the same \concept{\#} prefix. Some headings have structural meaning and are reserved words: \concept{\#Binding}, \concept{\#Tests}, \concept{\#Appendix}, and \concept{\#References}. 
Executable blocks are indicated by indentation, with the internal syntax of those blocks determined by the language binding. Types of executable blocks are distinguished using \texttt{\char`\~}-prefixed keywords called sigils. Some sigils can also carry a label, as in \sigil{test} \texttt{mcmxciv}.  A group of sigils declare and configure language binding:  \sigil{language}, \sigil{external}, \sigil{on-build}, and \sigil{keep-generated-src}. Another group of sigils are used for tests: \sigil{property}, \sigil{example}, and \sigil{test}. \sigil{run} indicates an execution entry point. The grammar restricts certain sigils to certain gross sections of the file, such as \sigil{test} in the \concept{\#Tests} section.

\clearpage

The lexer uses an ambiguous regular grammar, disambiguated first by rule priority, then maximal munch for specific token classes. This allows prose fallback to be lowest priority. Rule priorities are listed in Table~\ref{tab:priority}.

\subsection{Semantics and Names}

The detail of execution is delegated to other programming languages, tools, and libraries, defined by a structure called a binding. The semantics of the notlob language itself are then about firstly, defining hooks into that binding, and secondly, describing the relationships between names, prose and executable code.

\begin{table}[b]
\begin{tabular}{rl}
  \textbf{Priority} & \textbf{Terminals} \\
  \hline
  20 & \synt{SEPARATOR}, \synt{TESTS\_HEAD}, \synt{BINDING\_HEAD}, \synt{REFERENCES\_HEAD}, \synt{APPENDIX\_HEAD} \\
  10 & \synt{MOD\_HEAD}, \synt{SUBHEAD}, \synt{SIGIL}, \synt{TEST\_SIGIL} \\
  8  & \synt{INDENTED\_LINE}, \synt{BLANK}, \synt{BULLET} \\
  5  & \synt{REF} \\
  1  & \synt{LINE\_START\_TEXT}, \synt{PROSE\_TEXT} (fallback) \\
\end{tabular}
\caption{Notlob terminal priorities.}
\label{tab:priority}
\end{table}

\subsubsection{Bindings}

A binding defines a programming language, unit test library, property test library, and linter. For example, the Haskell binding uses stack-ghc, hunit, quickcheck, and hlint. This defines the context into which code blocks are assembled, with \sigil{property} blocks becoming property tests, but \sigil{example} and \sigil{test} blocks becoming unit tests. All code blocks within the \concept{\#Tests} section are interpreted as unit tests, under the scope of either the \sigil{test} label, or the most recent heading. Two separate sigils are provided for tests. Examples declared with \sigil{example}, like Python doctests~\cite{pajankar_python_unit_2017}, are considered part of the exposition of concepts, and restricted to the main body. Unit tests are considered supplementary facts, and restricted to the supplemental material. A declared property is treated as a formal invariant, and expected in the main section near prose explaining the concepts and motivation for the constraint.

As an implementation convention, the file \texttt{binding.lob} declares the binding for a project, and does not need to be imported by other source files.

\subsubsection{Traversing Program Semantics}

Names are first class objects in notlob. Titles and headings are part of the syntax, and available in the parse tree. Name and concept maintenance are important infrastructural burdens in large software systems. We consider names load bearing structures for holding theory with source code artifacts across people, machines, and time. We are inspired in this by craft observations on the importance of names, such as Michael Feathers' observation that ``Rename Class is the most powerful refactoring"~\cite{feathers_working_2004,feathers_systemmetaphor_2006}, as well as broad projects of making conceptual structure legible, such as the semantic web~\cite{lassila_semantic_2001,hitzler_review_2021}. 

As commentary is not discarded by the compiler, it is straightforward to use the parse tree to build a graph of names, prose, concepts and code that can be accessed by tooling. There are then two types of adjacency available cheaply to human and machine agents: file-adjacency, where elements are close together in the same chunk of serial text on the filesystem, and concept-adjacency, where related elements are connected on a path with a small number of edges. The name-graph can be explored by human and machine programmers at build time with accompanying tooling. It can also be used for consistency checks across the codebase, which includes the prose. The name-graph is an analogue to Knuth's generated index~\cite{knuth_literate_1984}, and many documentation tools since, but with a non-linear data structure.


\section{Implementation}
\label{sec:impl}

\begin{table}[htbp]
\centering
\caption{Command Overview for the Notlob Toolchain.}
\label{tab:notlob-commands}
\begin{tabular}{l l}
\textbf{Command} & \textbf{Description} \\ \hline
\texttt{run} & Assemble and execute a \texttt{.lob} file \\ 
\texttt{test} & Run all claims in a \texttt{.lob} file, or the whole project \\ 
\texttt{build} & Assemble a \texttt{.lob} file (or the whole project) to source artifacts \\ 
\texttt{weave} & Render a \texttt{.lob} file (or the whole project) as Markdown \\ 
\texttt{graph} & Export the package name-graph (JSON or Turtle RDF) \\ 
\texttt{query} & Query the package name-graph \\ 
\texttt{check} & Run semantic checks on the project name-graph \\ 
\texttt{init} & Initialise a new notlob project in the current directory \\ 
\texttt{new} & Create a new \texttt{.lob} module \\ 
\texttt{docs} & Write the language reference to \texttt{notlob-docs/} \\ 
\texttt{mcp} & Start the MCP tool server (stdin/stdout) \\ \hline
\end{tabular}
\end{table}

These design ideas have been implemented in an open source Python project\footnote{Source code and packages available at \url{https://github.com/adamburkegh/notlob}.}. A Lark grammar~\cite{shinan_lark_2026} is used as the basis of the parser.

Bindings have been implemented for Haskell, Python, and TypeScript. Sigils allow for working with external tooling and the executable layer, particularly \sigil{external} for referencing external sources without binding support, \sigil{on-build} as a hook for external invocation, and \sigil{keep-generated-src} for retaining generated source files in the target executable language.

Notlob source files end with a \texttt{.lob} suffix. Files are laid out on the file system using a deterministic rule where each word in the file heading is a new subdirectory. So a file headed \concept{\#Roman Numerals} is expected to be found in \texttt{roman/numerals.lob}, and this is enforced by the compiler.

Command line tools have been provided to work with the notlob language, as summarised in Table~\ref{tab:notlob-commands}. This includes standard build time tooling to compile, resolve dependencies, and run tests, as in tools such as \texttt{maven} or \texttt{stack}. The \texttt{weave} command is a literate programming-inspired mechanism for rendering the project as Markdown. In working with notlob, the author did not primarily work through separate documentation and code artifacts, as with Knuth's WEB~\cite{knuth_literate_1984}, but rather with the \texttt{.lob} source files themselves. However documentation output still has its uses for other forms of publication and distribution. 

The name-graph can be exported using \texttt{graph} or queried with \texttt{query}. This also includes export as RDF for possible use in semantic web tools~\cite{lassila_semantic_2001}. Deterministic checks for semantic consistency are executed by the \texttt{check} command. Missing imports and unused references are treated as errors. Possible typos (similar names off by 1-2 characters), violated conventions, and other style checks are provided as information-only warnings. The name-graph is intended for use as structured data input to LLMs to suggest semantic inconsistencies across prose and code elements of the codebase. The \texttt{query} command can also be used for graph-based navigation by machine agents, similar to keyboard shortcuts to jump to type declarations or function callers in an IDE.

The \texttt{init} command creates a new basic notlob project with instructions for coding agents on how to use notlob. The same set of commands can also be accessed via an MCP server.

The notlob implementation and the example projects were built with the assistance of Claude Code, using the Sonnet and Opus models.

\section{Programs and Experiences}
\label{sec:programs}

The following section describes programs developed in notlob.

\subsection{Roman Numerals}

\begin{figure}[t]
\begin{minipage}[t]{0.48\textwidth}
\begin{lstlisting}[language=notlob,basicstyle=\ttfamily\footnotesize]
#Roman Numerals
\end{lstlisting}
\begin{lobprose}
Convert integers to Roman numeral strings.  The numeral table maps each
milestone value to its symbol.  Conversion is greedy: find the largest
milestone that fits, append its symbol, subtract its value, repeat.
\end{lobprose}
\begin{lstlisting}[language=haskell,basicstyle=\ttfamily\footnotesize]
    numerals :: [(Int, String)]
    numerals =
        [ (1000, "M"), (900, "CM"), (500, "D"), 
          (400, "CD"), (100,  "C"), (90,  "XC"), 
          (50,  "L"), (40,  "XL"), (10,   "X"), 
          (9,   "IX"), (5,   "V"), (4,   "IV"), 
          (1,    "I")
        ]

    toRoman :: Int -> String
    toRoman 0 = ""
    toRoman n = snd h ++ toRoman (n - fst h)
      where h = head $ filter ((<=n) . fst) numerals    
\end{lstlisting}
\begin{lstlisting}[language=notlob,basicstyle=\ttfamily\footnotesize]
~example
\end{lstlisting}
\begin{lstlisting}[language=haskell,basicstyle=\ttfamily\footnotesize]
    toRoman 1    == "I"
    toRoman 6    == "VI"
    toRoman 1994 == "MCMXCIV"
\end{lstlisting}
\begin{lstlisting}[language=notlob,basicstyle=\ttfamily\footnotesize]
##Positive and Non-Empty
\end{lstlisting}
\begin{lobprose}
The length of the result is always positive for positive inputs, and
toRoman never returns an empty string for a positive integer.
\end{lobprose}
\begin{lstlisting}[language=notlob,basicstyle=\ttfamily\footnotesize]
~property
\end{lstlisting}
\begin{lstlisting}[language=haskell,basicstyle=\ttfamily\footnotesize]
    prop_positive :: Int -> Bool
    prop_positive n =
        let m = abs n `mod` 4000 + 1
        in not (null (toRoman m))
\end{lstlisting}
\end{minipage}
\hfill
\begin{minipage}[t]{0.48\textwidth}
\begin{lstlisting}[language=notlob,basicstyle=\ttfamily\footnotesize]
---

#Tests

##basic
\end{lstlisting}
\begin{lstlisting}[language=haskell,basicstyle=\ttfamily\footnotesize]
    toRoman 0    == ""
    toRoman 1    == "I"
    toRoman 5    == "V"
    toRoman 10   == "X"
    toRoman 50   == "L"
    toRoman 100  == "C"
    toRoman 500  == "D"
    toRoman 1000 == "M"
\end{lstlisting}
\begin{lstlisting}[language=notlob,basicstyle=\ttfamily\footnotesize]
##subtractive
\end{lstlisting}
\begin{lstlisting}[language=haskell,basicstyle=\ttfamily\footnotesize]
    toRoman 4    == "IV"
    toRoman 9    == "IX"
    toRoman 40   == "XL"
    toRoman 90   == "XC"
    toRoman 400  == "CD"
    toRoman 900  == "CM"
\end{lstlisting}
\begin{lstlisting}[language=notlob,basicstyle=\ttfamily\footnotesize]
##compound
\end{lstlisting}
\begin{lstlisting}[language=haskell,basicstyle=\ttfamily\footnotesize]
    toRoman 2024 == "MMXXIV"
    toRoman 3999 == "MMMCMXCIX"
\end{lstlisting}
\begin{lobprose}    
The 1994 case stacks all three subtractive forms at once (CM, XC, IV),
so it's worth naming on its own.
\end{lobprose}
\begin{lstlisting}[language=notlob,basicstyle=\ttfamily\footnotesize]
~test mcmxciv
\end{lstlisting}
\begin{lstlisting}[language=haskell,basicstyle=\ttfamily\footnotesize]
    toRoman 1994 == "MCMXCIV"
\end{lstlisting}
\end{minipage}
\caption{A notlob program for Roman numerals, using the Haskell binding.}
\label{fig:roman}
\Description{Code listing.}
\end{figure}

Roman Numerals is a small program to parse and translate Arabic numerals to Roman. A listing of the main program is in Figure~\ref{fig:roman}. This example illustrates the heading and subheading syntax, the use of \sigil{example} and \sigil{property} to support the description, and the macro-structure of main program and supporting facts\footnote{This and other examples can be found in the notlob project itself under \url{https://github.com/adamburkegh/notlob/tree/main/examples}.}. An excerpt of the name-graph is shown in Figure~\ref{fig:roman-graph} using the output of \texttt{notlob graph}. Nodes for modules, symbols and tests can be seen, as well as edges for definitions, uses and tests. Nodes reference their location in source files to allow precise navigation among related elements of code. A unique address is generated for each node from its location on the file system and within source files.


\begin{figure}
    \centering
    \begin{lstlisting}[language=JSON,basicstyle=\ttfamily\footnotesize]
{
  "nodes": [
    {
      "address": "roman",
      "label": "Roman",
      "kind": "MODULE",
      "start_line": 1
    },
    {
      "address": "roman/numerals/app",
      "label": "Roman Numerals App",
      "kind": "MODULE",
      "start_line": 1
    },
    {
      "address": "roman/numerals/app#main",
      "label": "main",
      "kind": "SYMBOL",
      "start_line": 6
    },
    {
      "address": "roman/numerals/app#example#1",
      "label": "example#1",
      "kind": "EXAMPLE",
      "start_line": 9
    },
...
{
      "address": "roman/numerals#Positive Length Natural Numbers",
      "label": "Positive Length Natural Numbers",
      "kind": "SUBHEADING",
      "start_line": 29
    },
    {
      "address": "roman/numerals#Tests#basic",
      "label": "basic",
      "kind": "TEST",
      "start_line": 45
    },
...
  "edges": [
...
    {
      "source": "roman/numerals#Positive Length Natural Numbers#property#1",
      "target": "roman/numerals#Positive Length Natural Numbers#property#1#prop_positive",
      "kind": "DEFINES"
    },
    {
      "source": "roman/numerals/app",
      "target": "roman/numerals",
      "kind": "IMPORTS",
      "start_line": 16
    },
    {
      "source": "roman/numerals/app#main",
      "target": "roman/numerals#toRoman",
      "kind": "USES",
      "start_line": 7
    },
...
}
    \end{lstlisting}    
    \caption{Excerpt of the output of \texttt{notlob graph}, showing part of the name-graph for this the Roman Numerals program in Figure~\ref{fig:roman}. }
    \label{fig:roman-graph}
    \Description{Code listing.}
\end{figure}

\subsection{Petri Net Chomper}

Petri Net Chomper is a web based game inspired by classic arcade games, but using Petri nets~\cite{bause_stochastic_2002} as maps. It was developed as a fun puzzle game and a teaching tool for those new to Petri nets. The implementation uses the TypeScript binding. 

Chomper was developed using a new agent session as a case study on a small to medium sized project, and source of design feedback. It has ten source files divided across modules for Petri net data structures and behaviour, game play, rendering, and I/O. The file \texttt{overview.lob} is a 73-line prose overview of the project that is also a notlob source file, referring to the concepts and modules, and part of the name-graph itself.\footnote{Code for the project can be found at \url{https://github.com/adamburkegh/pn-chomper}.}

Chomper was useful both as an extended TypeScript example, and a source of design pressure to introduce and improve integration with external build and execution points. It was also notable that some development tooling built into platform coding agents, such as Claude Code, use optimisations that undermine the context sharing advantages of notlob file adjacency. For example, Claude will delegate to tools that execute \texttt{grep} commands with output restricted to a small number of lines. Prompts to take advantage of the name-graph and contextual reads mitigated this somewhat; it is also hardly fundamental to the architecture of coding agents, but rather a sign of development tools which are a work in progress.

\subsection{Pleiades}

Pleiades is a project to independently reproduce the calculations for dating the \textit{Midnight Poem} traditionally attributed to Sappho~\cite{cuntz_seasonal_2016}. It is implemented using the Python binding and third party libraries for astronomical and geographical calculations. This project started with an entirely human-authored specification and design stub. The specification consisted of a prose overview of the motivation and goal. The design was sketched using two function signatures with empty implementations, and two unit tests written as notlob \sigil{example} blocks. This code compiled with tests failing, following a Test Driven Development (TDD) style. The coding agent was then introduced and prompted to implement with reference material such as the source paper and notlob documentation. The session continued using interactive prompting where the agent implemented most code independently. Human editing for style and clarity helped keep the code and prose well-organised~\footnote{Code and commentary \url{https://github.com/adamburkegh/pleiades}, including tag v0.1 before introduction of the coding agent.}.

Code, prose and tests produced by the agent followed the design sketch out quite closely. The agent detected type signature inconsistency in the example code almost immediately, and asked for input to resolve. It suggested a new module for historical dates, as BCE dates are not supported by Python standard libraries, which it delivered in a tight literate style with test coverage. The agent also propagated unfortunate human spelling mistakes in the proper names Mytilene and Alcyone. Under-examined assumptions introduced in the original specification, such as the direction of Mt Olympus of Lesbos from Mytilene, were also propagated without being flagged. The implementation included agent-written code making idiomatic use of astronomical libraries originally unknown to the developer. 

A round of project review by agents resulted in increasingly fine-grained, sometimes pedantic, criticism, reminiscent of academic paper reviews and probably related to the research content. It also yielded a change to horizon calculation that impacted the calculated dates and may improve the result in the paper.

\subsection{Design Observations}

This section describes partial projects and cross-cutting concerns.

An experimental notlob port of an established Digital Signals Processing (DSP) project written in Rust resulted in the coding agent spontaneously adding property tests which surfaced a subtle bug in the original library. The bug related to instability in the clamping calculation that had not been caught by other testing. This project also required developing an experimental Rust binding. Work was done by another developer who had not otherwise contributed to the notlob codebase\footnote{See patches-dsp project developed by Dominic Fox \url{https://github.com/poetix/notlob/tree/rust-bindings/examples/patches-dsp}}.

As design guidance, we suggest that prose in notlob programs should mostly provide additional information the code does not, rather than be a redescription of the code itself. Prose can explain the motivation for a feature, design constraints, computational complexity, or related parts of the system that are not directly related through executable paths like function invocation. With the inclusion of heading references, such conceptual dependencies can also be made visible in the name-graph structure itself.

After the establishment of a running notlob implementation with a number of example programs, two Claude Code chat sessions were initiated to provide design criticism informed by the ideas of the project but independent of any particular codebase. The prompt used, in part, was:

\begin{quote}
I would like to engage you in the new role of notlob critic. Notlob is a new experimental literate programming environment. (URL)

As notlob critic, you would draw on software engineering and literary practice, a little high theory, and a deep knowledge of this new tool. You will write in the precise, well-crafted prose of a good longform book reviewer, perhaps working for the New Yorker or the London Review of Books. 

From time to time I will ask your opinion of notlob projects, or on the shape of notlob itself. Like good designers and artists, you should try and develop a point of view, and have opinions on the development on the tool, its usage, and comparable projects. As well as reviews of particular projects, you would eventually author a style guide. We can also maintain informal notes - you can suggest the best way to store those as artifacts.
\end{quote}

In the second critic session, the prompt was varied, to add:

\begin{quote}
As a diversity of opinions is useful to me, as an experiment, let's say you are going through a gonzo-influenced period, inspired by Hunter S Thompson, David Foster Wallace, and Patricia Lockwood. 
\end{quote}

The critic sessions were engaged to help hold the theory of notlob without being tied to the fine details of implementation. Both found issues and suggested improvements. In general, the gonzo critic session found more bugs, in both notlob language projects and notlob itself. This included errors of function and drift between prose and code. The critic sessions also suggested fixes in the target projects, such as the promotion of implicit principles mentioned in prose to declared checkable \sigil{property} blocks. We note this as an anecdotal observation rather than the result of a controlled experiment.

Across multiple agent-assisted coding projects we observed a tendency for agents to neglect the declarative artifacts that scaffold their work. In notlob itself, the Lark grammar file was created, but many keywords were handled in a Python parsing layer, until specific re-engineering lifted them back into the grammar. On another occasion, USES nodes were added in a way that limited the scope only to external artifacts, a scope-reducing literal interpretation that undermined the usefulness of the feature. In Chomper, no property tests were added, even while formal properties of Petri nets were asserted in prose and essential to the functioning of the game. In another project outside of notlob, executable workarounds grew around LinkML schemas established as metadata descriptors. The recurrence across four different artifact types suggests the dynamic is general rather than tool-specific. In all of these cases human inspection and intervention was necessary to retain these artifacts as load-bearing points of consistency in the design, in a dynamic that will be eerily familiar to senior developers and architects on teams of human software developers everywhere.

\section{Related Work}
\label{sec:related-work}

A number of language designs have been advanced for structuring interactions with LLMs. The Language Model Query Language (LMQL)~\cite{beurer-kellner_prompting_2023} provides an SQL-like declarative language for structuring LLMs query prompts. DSPy~\cite{khattab_dspy_2024} defines executable data pipelines for LLMs using function declarations that take and emit natural language.
SGLang~\cite{zheng_sglang_2024} structures programs that call into LLMs for their work, including LLM-aware language primitives like \texttt{gen}, \texttt{select} or \texttt{image}. APPL~\cite{dong_appl_2025} is also intended to structure prompts within programs, using a special Python function decorator. Functions so decorated have their comment used as a prompt. Each of these tools improves prompting with program syntax, but without the focus on code and documentation as durable artifacts used by human programmers and LLM coding agents in notlob.

Literate programming was conceived by Knuth~\cite{knuth_literate_1984,knuth_literate_1992} and the broader intellectual project has had a great influence in research and practice, with documentation tools a part of many language toolkits, as well as an influence on coding notebooks such as Jupyter~\cite{kluyver_jupyter_2016}. Comprehensive review of this impact is beyond this paper, though we will note that well-structured prose explanations are not a routine feature of program source code observed in the wild. Notlob sits closer to the syntactically spare style of Ramsey's noweb~\cite{ramsey_literate_1994} than Knuth's original WEB. Like noweb, notlob is language-independent across bindings, and the literate layer is orthogonal to the code language. More recently, LLMs have triggered what one paper calls a ``renaissance of literate programming"~\cite{zhang_renaissance_2024}. The connection to literate programming can be a process link, that is, disciplined use of LLMs to summarise and regenerate code at different units of structure~\cite{shi_natural_2025}. It is also linked in the context of new tools. 
Goldfish Scheme~\cite{zhang_renaissance_2024} is a dialect of Scheme used to structure precise natural language prompts with documentation, with a target of the development of large programs. The authors introduce the term ``Interoperable Literate Programming", to summarise three principles of literate programming: flexible code organisation, tangling (automated processing for a compilable source), and bidirectional workflow. They place Jupyter as a literate programming tool, but not ILP, due to the difficulty of integrating or exporting notebooks into a broader project structure. Using this taxonomy, notlob is an ILP project. Source files can be organised in familiar modular ways and automatically made into reusable executable artifacts. Experiments with Goldfish Scheme show smaller language models using literate programming can be more effective software developers than larger language models using conventional programming languages alone.

As well as literate programming, the notlob design is influenced by tools for Behaviour Driven Development (BDD)~\cite{north_introducing_2006,bisht_robot_2013}, and Domain Specific Languages~\cite{mernik_dsl_2005}. Behaviour Driven Development tools supply natural language-like use cases as code artifacts which are also executable functional tests. BDD existed before LLMs, so a natural language translation layer of some kind was always needed, reducing the use cases to a kind of domain pidgin, and in our experience exerting flattening pressure on the underlying test API. In notlob there is no translation layer, as using LLMs, well-crafted natural language and precise asserted facts and other executable quality mechanisms can reinforce each other in machine checkable common artifacts. Notlob also combines the importance of names, from BDD, with the web of name relationships made use of by ontologies and the semantic web~\cite{lassila_semantic_2001}.

Behaviour Driven Development tests, and notlob programs, are related to Domain Specific Languages (DSLs)~\cite{mernik_dsl_2005}, and the interleaved prose and executable fragments in notlob programs make it a kind of Embedded DSL~\cite{hudak_building_1996}. Notlob itself is not a DSL toolkit, in that it is not primarily a mechanism for introducing custom syntax for effective executable expression of domain problems. It is not so much a Domain Specific Language as a Domain Entangling Language, evolving a design approach from literate programming and Behaviour Driven Development. 

Our approach contrasts with the emerging tools and software methodologies of Specification Driven Development (SDD)~\cite{piskala_spec_2026}. SDD focuses on structured natural language specifications as primary artifacts maintained upstream of code generated by LLMs. Criticism of SDD~\cite{bockeler_sdd_2025} points out the impact of LLM non-determinism on generated code quality, as well as struggles with articulating intent at the most useful levels of abstraction. SDD reinvents waterfall~\cite{benington_production_1987} in LLM-accelerated form, but we doubt this will resolve waterfall's problems. SDD focuses on specification as the start of a pipeline. notlob co-locates specifications, code and tests at the same workbench, ready to be worked on together in frequent iterations by humans and machine agents. The specification is not upstream of code, but beside it. We follow the view of James Noble, that while natural language in programming presents engineering opportunities, 
 ``switching from a formally defined programming language to the implicit prompt language of an AI model, with zero specification, zero syntax, zero semantics, zero consistency across models and between releases, has zero appeal" ~\cite{noble_automatic_2024}.

\section{Conclusion and Future Work}
\label{sec:conclusion}

This paper presents a programming language and toolset to solve some problems of natural language-literate coding agents by using the techniques of literate programming. The design provides  conceptual co-location within files, and in a parse-tree based name-graph, with the intent that coding agents can make more effective use of the context window and necessary external memory artifacts. The language and toolset was implemented in Python, along with a number of example programs of small to medium size. Future work may explore bindings with multiple execution languages for ``full-stack" concept descriptions, experimental evaluation with zero- or one-shot prompts, notlob-aware coding agents, and the construction of larger systems.

\begin{acks}
Thanks to Phil Cook and Dominic Fox for their thoughts on this research.
\end{acks}

\bibliographystyle{ACM-Reference-Format}
\bibliography{notlob}

\end{document}